\documentstyle[12pt]{article}
\begin{document}
\title{Interactions of Collective Excitations with Vortices in Superfluid
Systems}
\author{E. Demircan$^1$ , P. Ao$^2$, and Q. Niu$^1$\\
      $^1$ Department of Physics\\
      The University of Texas at Austin \\
      Austin, TX 78712 \\
      $^2$ Department of Physics, FM-15 \\
      University of Washington, Seattle, WA 98195 }
\maketitle
\newcommand{\beq}{\begin{equation}}
\newcommand{\eeq}{\end{equation}}
\newcommand{\vq}{\vec{q}}
\newcommand{\beqa}{\begin{eqnarray}}
\newcommand{\eeqa}{\end{eqnarray}}
\newcommand{\x}{\vec{x}}
\newcommand{\r}{\vec{r}}
\newcommand{\ro}{\rho_{0}}
\newcommand{\rx}{\rho( \x )}
\newcommand{\sx}{ S(\x) }
\newcommand{\dr}{\delta \rho (\x)}
\newcommand{\drp}{\delta \rho (\x ')}
\newcommand{\dd}{\vec{\nabla}}
\newcommand{\vv}{\frac{\hat{\theta}}{r}}
\newcommand{\k}{\vec{k}}
\newcommand{\q}{\vec{q}}
\newcommand{\p}{\paragraph{}}

\begin{abstract}
 We investigate the interactions of collective excitations with
vortices in superfluid systems, including $~^4$He and superconductors.  The
dynamical equations are obtained  by the aid of the many-body
wavefunction and the density-density correlation function. The scattering
cross-section of collective excitations with a  vortex is calculated in the
Born
approximation (valid at long wavelengths), and is expressed in terms of the
collective excitation spectrum  $ \omega (q) $ by the simple and general
formula
$ \sigma =  \frac{\pi}{2} \frac{q}{[\omega'(q)]^2} \cot^2\frac{\theta}{2} $,
where $ q $ is the wave number of the excitation, and $\theta $ is the
scattering angle. At short wave lengths, the classical equations of motion
are derived.

\end{abstract}
\newpage
\section{Introduction}

The motion of vortices is considered to be the main source of dissipation in
superfluid systems. At finite temperatures, the existence of collective
excitations and their interactions with the vortex lines create the phenomenon
known as mutual friction. \cite{hillel,barenghi,vinen,donnelly}
The phenomenological parameters of mutual friction can be derived from the 2D
scattering cross-section of the collective excitations\cite{sonin,rayfield}.

\p

Different methods have
been used in the literature for the calculation of the scattering
cross-sections.
One method is based on the classical dynamics of phonons \cite{sergey} or
 rotons \cite{hillel,donnelly,sonin}
treated as particles.  In the presence of a vortex,
the Hamiltonian for such a particle is given, for instance,
by
$ H= \epsilon_0 (\vec{p}) + \vec V(r)
\cdot \vec{p} $
where $\epsilon_0 (\vec{p})$ is the energy spectrum of the collective
excitations, and the second term represents the Doppler shift of energy
due to the velocity field $\vec V(r)$ associated with the vortex.
Effects of the vortex core and density modulations have also been considered.
Another method is to treat such a particle quantum mechanically by introducing
a quantum wave function for it, using the operator version of the above
Hamiltonian to describe the dynamics.  This method has been applied to the
roton case, with the cross-section calculated in the Born approximation
\cite{hall}.
There is then the more conventional method based on the hydrodynamic wave
equations.
Born approximation \cite{pita,sonin} and phase shift analysis \cite{fetter}
have been used for the calculation of phonon scattering cross-sections.

\p

We observed that the phenomenon can be considered in more general terms.
Besides
$~^4$He, there are other systems, such as superconductors, quantum Hall systems
\cite{prange}, quantum spin systems, that support vortices and collective
excitations with different microscopic structure and ground state properties.
The purpose
of this paper is to establish the relation between the different methods, and
to  present a calculation of the scattering cross-section that does not refer
to
the details of the microscopic structure of the systems. For long wavelength
scatterings, the cross-section may be calculated using the Born approximation,
and is found to be given by the simple and general expression:
\beq
    \sigma(q,\theta) = \frac{\pi}{2} \frac{q}{\left[\omega'(q)\right]^2 }
                       \cot^2 \frac{\theta}{2},
\eeq where $ q $ is the momentum, $ \omega $ is the excitation spectrum, and $
\theta $ is the scattering angle of the excitations.
We also show that a classical interpretation of the scattering is possible
in the opposite limit, i.e. at short wavelengths.

\p

In Section 2, we will give the basic ingredients of our method. In Section 3,
we
will calculate the cross-section in the Born approximation, and present the
classical approach. In Section 4, we will apply the results we have derived on
some systems. In Section 5, we present our conclusions.

\section{Dynamical Equations}

In this section we will lay out the general formulation of the problem, and
obtain the dynamical equations for the elementary  excitations in the presence
of a vortex. The core of the idea is to obtain an Hamiltonian for the
elementary
excitations, and use the Hamilton's equations to describe the motion of the
excitations.

\subsection{General Formulation}

The elementary excitations are  represented by small variations in the local
density and phase, and are described by the following many-body wave function
\cite{Feynman,thouless}
\beq
   \Psi = \exp \left(
               { \sum_{i=1}^{N} \alpha ( \vec{x}_i) + iS(\vec{x}_i)}
               \right) \Psi_0 ,
\eeq where $\Psi_{0}$ is the ground state wave function, $S$ and $\alpha$ are
arbitrary functions that satisfy $\alpha({\x}) \ll 1$, and do not vary rapidly
over the characteristic length scales of the system, such as the interparticle
spacing in $~^4$He and the coherence length $\xi_{0}$ in superconductors.

\p

We follow the lines in \cite{np} to obtain the energy in this state
as\footnote{we use $\hbar = m = 1$ unless otherwise stated}:
\beq
   E-E_{0}= \frac{1}{2} \int \rho (\vec{x}) \left\{
                             \left| \vec{\nabla} S(\vec{x}) \right|^2
                            +\left| \vec{\nabla} \alpha(\vec{x}) \right|^2
                                            \right\} d\vec{x}, \label{E}
\eeq where the integration is over the spatial dimensions of the system, and
$\rho(\x )$ is the density:

\beqa
   \rho(\x ) & = & \left< \Psi \left|\hat{\rho}(\x )\right| \Psi \right>
\nonumber \\
             & = &        \left< \Psi_{0} \right|
                       \hat{\rho} (\vec{x})
                       \exp{\left( 2 \int\hat{\rho} (\x ') \alpha (\x ')
                            d\x ' \right) }
                       \left| \Psi_{0} \right>,
\eeqa with
$  \hat{\rho}(\x ) =  \sum_{i}\delta (\x  -\vec{x_i} )$.

\p

 We want to express eq.~(\ref{E}) in terms of the canonically conjugate
variables $\delta\rho = \rho - \ro $ and $ S $ \cite{landau}, and regard it as
the Hamiltonian of the collective excitations. Eq.~(\ref{E}) is exact, but we
have to make an approximation in order to make progress. Thus we will linearize
the density in $ \alpha $ using the assumption that the fluctuations are small,
yielding

\beqa
    \rho(\x)  & = & \rho_{0} + 2 \int \left< \Psi_{0} \right|
                                 \hat{\rho}(\x) \hat{\rho}( \x ')
                                 \left| \Psi_{0} \right>
                               \alpha( \x ' )  d \x \\
              & = & \rho_{0} + 2 \rho_{0} \int g(\x -\x ' ) \alpha(\x ') d\x ',
\eeqa where $\ro$ is the density of the system in the ground state, and the
function
$ g ( \x - \x ') $ is the density-density correlation  \cite{Feynman}. We can
invert this relation  and solve for $\alpha ( \x) $
\beq
    \alpha( \x ) = \frac{1}{2 \ro }
                   \int \delta \rho ( \x ') g^{-1} (\x -\x') d\x
',\label{alpha}
\eeq with the definition of the inverse function by
\beq
    \int g^{-1} ( \x -\x') g(\x ') d\x' = \delta ( \x ) ,\label{g}
\eeq

Relation~(\ref{alpha}) enables us to write the energy function eq.~(\ref{E}) in
terms of the variables $\delta
\rho (\x) $, $ S ( \x)$ and the correlation function $ g (\x) $. The properties
of the ground state and the interactions between particles are included in $
g(\x) $. The normal modes of these equations will be the excitation spectrum of
the fluctuations.

\subsection{Collective Excitations in the Absence of a Vortex}

After substituting eq.~(\ref{alpha}), the Hamiltonian of the system without any
vortex in it can be written as
\beqa
 H & = & \frac{1}{2}
             \ro \int
                  \left| \dd S(\x) \right|^2 d\x
                 \nonumber \\
      &   &      +\frac{1}{2}  \ro \int
                        \left[ \dd
                              \left(
                            \frac{1}{2\ro} \int \drp g^{-1}(\x - \x') d\x '
                              \right)
                        \right]^2
                      d\x .
\eeqa The first term can be considered as the kinetic energy and the second one
as the potential energy of fluctuations.  The Hamilton's equations are just
functional derivatives of the Hamiltonian :

\beqa
    \frac{d\, \delta \rho (\x ,t)}{d\, t} & = & \frac{\partial H}{\partial S(\x
,t)}
\nonumber \\
                                          & = & - \ro \nabla ^2 S(\x , t)  \\
   -\frac{d \, S(\x, t)}{d \, t}          & = & \frac{\partial H}{\partial
\delta\rho(\x,t) }   \nonumber \\
                                          & = & -\frac{1}{4\ro } \nabla ^2 \int
                                   \int
                                   g^{-1}(\x - \x ' ) g^{-1} (\x' - \x'')
\delta
                                   \rho (\x'')
                                   d\x'\: d\x'',
\eeqa where we have used the fact that for an isotropic system $ g(\x) $
depends
only on
$|\x|$, and omitted some boundary terms. If we recall that  $\vec{\nabla} S $
is
the velocity field, we can easily recognize the first equation as the
continuity
equation, and the second one as the generalized Euler equation of hydrodynamics
in a form that also takes into account of density correlations in the system.

\p

We can try a plane wave solution to these equations
\beqa
   S(\x , t) & = & A e^{i( \k \cdot \x - \omega t)} \nonumber \\
   \delta\rho(\x , t) & = & B e^{i(\k \cdot \x - \omega t)} ,
\eeqa which immediately requires the conditions
\beqa
    -i B \omega  & = & A \ro k^2  \\ \label{12}
    \nonumber \\
     i A \omega  & = &  B \frac{k^2}{4 \ro }
                     \left[
                           \int g^{-1} (\x) e^{i \k \cdot \x} d\x
                     \right]^2 \nonumber \\
                 & = &  B \frac{k^2}{4 \ro {\cal S}^2 (k) }
\eeqa where ${\cal S}(k) $ is the structure factor defined as the Fourier
transform of $ g(\x)$. A non trivial solution to these conditions is
\beq
    \omega = \frac{k^2}{2 {\cal S}(k)},
\eeq  with
\beq
    B = 2i \ro {\cal S} (k) A .
\eeq The first relation is the famous Bijl-Feynman formula for the collective
excitation spectrum \cite{Feynman}, and the second gives the relation between
the phase and amplitude fluctuations.

\subsection{Effective Hamiltonian in the Presence of a Vortex}

Thus we have shown that eq.~(\ref{E}), when used as the Hamiltonian of the
collective excitations, gives us the correct spectrum. Now we want to put a
vortex into the system and obtain an effective Hamiltonian for the long
wavelength excitations. For simplicity we assume a single, straight, fixed
vortex line at the origin. The many body wave function in eq.(1) can now be
written as:
\beq
   \Psi = \exp \left(
               { \sum_{i=1}^{N} \alpha ( \vec{x}_i) + iS(\vec{x}_i)}
               \right)
              \prod_{j=1}^{N} f\left( |\x_{j}|\right)
               e^{i\theta_{j}}
               \Psi_{0} ,
\eeq where $ f $ is a function that becomes 1 outside the vortex core but goes
to zero at the origin. $ \theta_{j}$ is the polar angle at the position of
particle $ j $. A full treatment of the problem should take into account the
core structure through the function $ f $. However we will make the further
simplification of taking $ f $ equal to 1 everywhere, thus disregard the
effects
of the vortex core.  Then the effect of the vortex is to simply cause a shift
in
the canonical phase variable by $ \theta $
\beq
    \dd S \rightarrow \dd S + \vv , \label{sshift}
\eeq where $\theta $ is the polar angle and $ r$ is the distance to the vortex
line. Under these circumstances, the Hamiltonian up to second order terms in
$\dr$ and $ \vec{\nabla}S(\x) $ is
\beqa
    H & = & \frac{1}{2}
             \ro \int
                  \left| \dd S(\x) \right|^2 d\x
               \nonumber \\
      &   &     +\int      \dr \vv \cdot \dd S( \x) d\x
                 \nonumber \\
      &   &      +\frac{1}{2}  \ro \int
                        \left[ \dd
                              \left(
                            \frac{1}{2\ro} \int \drp g^{-1}(\x - \x') d\x '
                              \right)
                        \right]^2
                      d\x , \label{vhamiltonian}
\eeqa where we have ignored some  boundary terms and terms second order in the
vortex field. As discussed above, the first and last terms describe the
collective excitations. The scattering of excitations is caused by the second
term which, in classical terms, looks like a Doppler shift in the energy. The
correlation function $ g(\x) $ is unchanged by the inclusion of the vortex as
the function $ f $ representing the vortex core was taken to be 1 everywhere.
For the same reason the vortex couples to the kinetic term but not the
potential
energy term. This is why the vortex interacts with the excitations in quite a
universal way, as a result of which a simple and general expression for the
scattering cross-section will be obtained in the next chapter.

\section{Calculation of the Scattering Amplitude}

Now we can analyze Hamilton's equations, and calculate the scattering
amplitudes.  The Hamilton's equations in this case differ from the previous
ones
only by the appearance of the vortex terms

\beqa
    \frac{d\, \delta \rho(\x,t)}{d\, t}   & = & - \ro \nabla ^2 S(\x , t) - \vv
\cdot \dd
\delta\rho(\x, t)  \label{20} \\
   -\frac{d\, S(\x, t)}{d\, t}            & = & \vv \cdot \dd S(\x , t)
\nonumber \\
                                          &   & \mbox{} - \frac{1}{4\ro }
\nabla
^2 \int
                                   \int
                                   g^{-1}(\x - \x ' ) g^{-1} (\x' - \x'')
\delta
                                   \rho (\x'')
                                   d\x'\: d\x'' . \label{21}
\eeqa

The time dependence of the solutions can still be factored by $ e^{-i \omega t}
$, then the equations read
\beqa
    -i \omega \delta \rho (\x) & = & - \ro \nabla ^2 S(\x) - \vv \cdot
                                       \dd \delta\rho(\x)  \\
                i \omega S(\x) & = & \vv \cdot \dd S(\x) \nonumber \\
                               &   & \mbox{} + \frac{1}{4\ro } \nabla ^2 \int
                                     \int
                                     g^{-1}(\x - \x ') g^{-1} (\x'-\x'')\delta
                                     \rho (\x'')
                                     d\x'\: d\x''.
\eeqa

The vortex terms have position dependent coefficients and prevent an exact
solution to be obtained. We will use a Born approximation, in which we will
assume that we can replace the $ S(\x) $ and $ \dr $ in the vortex terms by the
free solutions obtained previously:
\beqa
   S_0(\x)  & = & A e^{ i\q \cdot \x } \nonumber \\
   \delta\rho_0(\x) & = & B e^{i\q \cdot \x} ,
\eeqa with the same dispersion relation. We will also restrict ourselves to two
dimensions in the plane perpendicular to the vortex line, since there is no
scattering parallel to the line. The dynamical equations now contain
inhomogeneous terms
\beq
    -i \omega \delta \rho(\x) + \ro \nabla^2 S(\x)  =  -i B \vv \cdot \q e^{i
\q
        \cdot \x} ,
\eeq
\beq
     i \omega S(\x) +  \frac{1}{4\ro } \nabla ^2 \int
                                     \int
                                     g^{-1}(\x - \x ') g^{-1} (\x'-\x'')\delta
                                     \rho (\x'')
                                     d\x'\: d\x''  = i A \vv \cdot \q  e^{i \q
        \cdot \x} .
\eeq By taking Fourier transform of both sides we turn the coupled equations
into a linear matrix equation
\beq
   \left[
        \begin{array}{cc}
           i \omega & \frac{k^2}{4 {\cal S}^2 (k) \ro } \\
                    &  \\
           -k^2 \ro & i \omega
        \end{array}
   \right]
   \left[
         \begin{array}{c}
            \tilde{S}(\k ) \\
       \\
            \tilde{\delta \rho } (\k)
         \end{array}
   \right]
   =
     - {\cal F}\left\{ i \vv \cdot \q e^{i \q \cdot \x} \right\}
   \left[
        \begin{array}{c}
         A \\
       \\
         B
        \end{array}
   \right] ,  \label{2}
\eeq where a tilde denotes the Fourier transformed quantities. It is simple
algebra to perform the necessary Fourier integral on the right hand side of the
equation. The result is
\beq
   {\cal F}\left\{ i \vv \cdot \q e^{i \q \cdot \x} \right\} =
    2 \pi\frac{\hat{z} \cdot ( \q \times \k)}{ \left| \q - \k \right|^2 } ,
\eeq  where $ \hat{z} $ is the unit vector in the direction of the circulation.
We invert eq.~(\ref{2}), replace $ {\cal S}(q) $ by $ q^2 / [2\omega(q)] $, and
then obtain the Fourier transform of the scattering amplitudes

\beqa
 \lefteqn{  \left[
         \begin{array}{c}
            \tilde{S}(\k) \\
       \\
            \tilde{\delta \rho } (\k )
         \end{array}
   \right] = }  \nonumber \\
          & & \frac {A 2 \pi }
           { - \omega ^2(k) +  \omega^2(q)}
     \frac{\hat{z} \cdot ( \q \times \k) }
          {\left| \q - \k \right|^2}
   \left[
          \begin{array}{cc}
           i \omega & - \frac{ \omega^2 (k)}{ \ro k^2} \\
       \\
           \ro k^2  & i \omega
          \end{array}
   \right]
   \left[
       \begin{array}{c}
           1 \\
       \\
            i \ro \frac{q^2}{\omega ( q)}
       \end{array}
   \right] .
\eeqa

The transformation back to normal space variables is given by
\beq
   \left[
          \begin{array}{c}
             S( \x) \\
          \\
             \delta \rho ( \x)
          \end{array}
   \right]
    =
           \int e^{ i  \k \cdot \x   }
   \left[
          \begin{array}{c}
           \tilde{S}(\k) \\
       \\
           \tilde{\delta \rho}(\k)
          \end{array}
   \right]
          \frac{d^2 k}{ (2 \pi )^2 } .
\eeq The integrals cannot be evaluated exactly. Since our purpose is to
calculate the cross-section we only need the amplitudes in the radiation zone,
$k \, r \gg 1 $. A separation of the $ \k $ integral over  components, $
k_{\parallel} $ and $ k_{\perp} $, parallel and perpendicular components of $
\k
$ to $ \x $ respectively, is helpful.  Then specifically for $ S ( \x ) $ we
get:
\beqa
    S( \x, t) & = &  A i e^{ - i \omega t } \frac{2 \pi }{(2 \pi)^2}
                     \int \frac{ dk_{\parallel} \, dk_{\perp}}
                               {\omega (k) - \omega (q)}
                     \frac{ e^{ i k_{\parallel} r }}{ \omega (k) +
                                                      \omega (q) }
\nonumber \\
              &   & \times \frac{ \omega^2 (k) q^2 + \omega^2 (q) k^2}
                          { \omega (q) k^2}
                    \times \frac{\hat{z} \cdot (\q \times \k )}
                          {\left| \q - \k \right|^2}. \label{31}
\eeqa

We will first keep $ k_{\perp }$ constant, and evaluate the $ k_{\parallel}$
 integration. We can take care of the boundary conditions at infinity by
supplying a small imaginary part wherever necessary. The boundary conditions
require that the outgoing scattering amplitudes should behave as $ e^{i k r} /
\sqrt{r}$, so the relevant poles are at the values of $ k_{\parallel 0} =
\sqrt{K^2 -k_{\perp}^2} + i \eta $, where $ K $ satisfies

\beq
   \omega (q) = \omega (K), \label{pole}
\eeq and $ \eta $ is a positive infinitesimal constant.

We have to note that $ k $ is not necessarily equal to $ q $, as might be
expected from energy conservation, since the spectrum may exhibit a behavior
such that several $ k $ values can correspond to the same energy. A very well
known example is the roton spectrum, which we will investigate in the next
section. For the time being we will work with the case $ K = q $.

\p

Expanding around the pole to linear order in $ (k_{\parallel} -k_{\parallel 0})
$ gives
\beq
    \omega (k) \approx \omega (q) + \omega '(q)
                        \frac{dk}{dk_{\parallel}}
                            \left|_{k_{\parallel 0}}
                            \right.
                        ( k_{\parallel} -k_{\parallel 0} ) .  \label{approx}
\eeq Then we place it into the integral and obtain the simple result
\beq
    S(\x,t) = - A e^{-i \omega t}
                      \hbox{Res(q)} , \label{S}
\eeq with
\beqa
  \hbox{ Res} & = & -\int dk_{\perp} e^{ i k_{\parallel 0} r}
              \frac{1}{
                           \omega '(q) \frac{dk}{dk_{\parallel}}
                           \left| _{k_{\parallel 0}}
                           \right.
                       }   \nonumber \\
              &   & \times \frac{
                           q( k_{\parallel 0} \sin \theta - k_{\perp}
                           \cos \theta)
                          }
                          {
                           k_{\parallel 0}^2 + k_{\perp}^2 + q^2 -
                          2qk_{\parallel 0}\cos \theta - 2qk_{\perp}
                           \sin \theta
                          } ,   \label{res}
\eeqa and keeping in mind that $ k_{\parallel 0} =   \sqrt{ q^2 - k_{\perp}^2}
$.
Now we have to evaluate the integral over $ k_{\perp} $. In the radiation zone
where we take $ r \rightarrow \infty $, the integral will be dominated by the
stationary point of the exponential, which is located at $ k_{\perp} = 0 $.
Assuming that the integrand other than the exponential behaves smoothly, we can
set $ k_{\perp} = 0 $ everywhere except in the exponential. A stationary phase
approximation gives
\beq
   \mbox{Res}= - e^{ -i\frac{\pi}{4}}
                 \sqrt{\frac{\pi}{2}}
                 \frac{e^{i q r }}{\sqrt{r}}
                 \frac{\sqrt{q} }{\omega '(q) }
                 \frac{\sin \theta}{1 - \cos \theta} .
\eeq Combining with eq.~(\ref{S}) we get the desired result
\beq
  S(\x,t) = A e^{i \omega t} e^{-i\frac{\pi}{4}}  \frac{ e^{i q r}}{\sqrt{r}}
               \sqrt{\frac{\pi}{2}}
               \frac{\sqrt{q}}{ \omega '(q)}
               \frac{ \sin \theta}{ 1 -  \cos \theta} .
\eeq The cross-section is given by the square of the amplitude of the outgoing
wave
\beq
    \sigma ( q, \theta) = \frac{\pi}{2} \frac{q}{ \left[ \omega '(q) \right]
^2}
                            \cot^2\frac{\theta}{2} , \label{cs}
\eeq which is the central result of this article. Note that the angular
dependence is universal, which is symmetrical with respect to $ \theta
\rightarrow - \theta $, and diverges at small angles. The spectrum function
enters in a simple form, and only affects the $ q $ dependent factor.

\p

One can evaluate the integrals in the general case where there may be more than
one possible value of $ K $. A similar calculation leads to the scattering
amplitude given by the sum of residues satisfying eq.~(\ref{pole})

\beq
   S(\x,t) = A e^{-i \omega t} \sum_{K} e^{-i \frac{\pi}{4}}
               \frac{ e^{i K r}}{\sqrt{r}}
               \sqrt{\frac{\pi}{2}} \frac{q ( K^2 + q^2) }{\sqrt{K}}
               \frac{1}{ \omega '(K)}
               \frac{ \sin \theta}{ K^2 + q^2 - 2Kq \cos \theta} . \label{Sf}
\eeq
%{ put some explanation and discussion here} However we must note that one has
to pay more attention to the degenerate case when $ \omega '(K) =0 $. In that
case the idea is still the same, only we need to expand to quadratic order in
eq.~(\ref{approx}), which forces us to evaluate the derivative of the integrand
in eq.~(\ref{res}).

\p

Born approximation is valid if the scattered amplitude is small when compared
with the incoming amplitude. We can expect this to be violated most where the
potential is strongest, which is the origin in our case. A simple calculation
after setting $ \x = 0 $ in eq.(\ref{31}) leads to the requirement that

\beq
 \frac{q}{\omega'(q)} \ll 1 . \label{born}
\eeq Thus if the excitation spectrum vanishes with a power less than $ 2 $ as $
q \rightarrow 0 $ then Born approximation is valid in this limit.

\subsection{Classical Hamiltonian}

The other approach which is mostly used for rotons is the classical equations
of
motion for the collective excitations. It is possible to show that one can
obtain the classical Hamiltonian, thus the classical equations, from
eqs.(\ref{20}) and (\ref{21}). These equations are like wave equations for the
amplitudes $ S $ and $\delta \rho $. The classical equations follow from the
geometrical optics approximation or the WKB. We replace the amplitudes $ S $
and
$ \delta \rho $ in the eqs.(\ref{20}) and (\ref{21}) by
\beqa
    S(\x , t) & = & A e^{i\phi (\x, t) } \nonumber \\
    \delta \rho (\x , t) & = & B e^{ i \phi (\x , t)} ,
\eeqa where $ A $ and $ B $ in general have spatial variation, but in the
spirit
of the WKB approximation they can be taken as constant.  Then from
eqs.(\ref{20}) and (\ref{21}) we obtain
\beqa
    i \left[
            \frac{\partial \, \phi}{\partial \, t} + \vv \cdot \vec{\nabla}
\phi
      \right] B & = & \ro \left| \vec{\nabla} \phi \right|^2 A \\
   -i \left[
            \frac{\partial \, \phi}{\partial \, t} + \vv \cdot \vec{\nabla}
\phi
      \right] A & = & \frac{ \left| \vec{\nabla} \phi
                             \right|^2
                           }
                           { 4 \ro {\cal S}^2
                             (
                                         \vec{\nabla} \phi
                             )
                           } B .
\eeqa We want to remind that we are stil using units $ \hbar = m = 1 $. The
momentum and energy of classical particles corresponding to wave packets of the
amplitudes are given by
\beqa
     E = -\frac{\partial \, \phi}{\partial \, t} & \mbox{and} &
     \vec{p} = \vec{\nabla}\phi .
\eeqa Thus we obtain the relation
\beq
   E = \omega (\vec{p})  + \vv \cdot \vec{p} ,
\eeq where we identified the energy spectrum of the excitations, $ \omega (
\vec{p} ) $  as in the Bijl-Feynman formula. This is obviously the classical
Hamiltonian for the collective excitations with the Doppler shift in the energy
due to the vortex. The equations of motion that follow from this Hamiltonian
have been used extensively in the literature for rotons
\cite{hillel,barenghi,sonin}, and recently for scattering of phonons from a
classical vortex \cite{sergey}.

\p

The WKB approximation is valid, if the spatial variation of $ \phi (\x,t) $
satisfies
\beq
 \nabla^2 \phi \ll |\vec{\nabla} \phi |^2 ,
\eeq
 For this approximation to be valid the wave packets have to satisfy
\beq
    \vec{\nabla} \cdot \vec{p} \ll \left| \vec{p} \right|^2,
\eeq or as an estimate on the left hand side we can use the uncertainties of
momentum and position
\beq
    \frac{\Delta p}{\Delta x} \ll \left| \vec{p} \right|^2.
\eeq We can take $ \Delta p \sim 1/ \Delta x $ and $ \Delta x \sim \xi $. The
classical approximation is valid if
\beq
    p \, \xi \gg  1.
\eeq Thus WKB is apropriate for large momentum. This is the opposite limit of
validity of the Born approximation given in
eq.~(\ref{born}).

\section{Applications} The results of the previous section are quite general in
the sense that as long as we know what the collective excitation spectrum of a
system is, we can find the scattering cross-section of these excitations from a
vortex. In this section we will show their use in superfluid helium and
superconductor thin films.
\subsection{Superfluid He}

 1. {\em Phonon Scattering} :  In the phonon part of the spectrum, the long
wavelength low energy part, the relation between $ k $ and $\omega $ is one to
one,  and is given by
\beq
 \omega (q) = s \, q ,
\eeq where $ s $ is the speed of sound. The only pole satisfying $ \omega (q) =
\omega( K ) $ is at $ K = q $. Thus from eq.~(\ref{cs}) we obtain:

\beq
   \sigma ( q, \theta) = \frac{\pi}{2} q
           \left( \frac{ \hbar}{m s}
           \right) ^2 \cot^2 {\frac{\theta}{2}} .
\eeq This result was obtained by Fetter \cite{fetter} using phase shift
analysis
of the scattered wave in the hydrodynamic equations. The other results in the
literature vary. Pitaevskii\cite{pita} and Sonin\cite{sonin} have different
results from the above although they both use Born approximation and similar
equations. However we have found that Pitaevskii's calculation suffers from an
algebraic error. In his paper eq.(25) does not follow from the substitution of
eq.(24) into eq.(19). Instead when correctly done the result is the same as
what
we have obtained above. The calculation of Sonin is a more complicated one
that
it takes into account of the curvature of the vortex line. His calculation must
be also incorrect, however, as it gives the result of Pittaevskii in the limit
of a straight vortex line.

\p

2. {\em Roton Scattering} : We can employ the formula (\ref{Sf}) to obtain the
cross-section of rotons. There are two values that satisfy eq.~(\ref{pole}).
The
spectrum   near the roton minimum can be written in the Landau form:
\beq
   \omega (q) = \Delta + \frac{(q - q_{0})^2}{2 \mu} ,
\eeq where $ \Delta $ denotes the energy  and $ q_0 $ denotes the wave vector
at
the roton minimum. The roton has two poles that satisfy
 $\omega (q) = \omega (K) $, which are
\beqa
   K_1 = q & \mbox{ and } & K_2 = 2q_0 - q .
\eeqa Now using eq.(\ref{Sf}) :
\beqa
  S(\x ,t) & = & A e^{ i \omega t} \sqrt{\frac{\pi}{2}}
            \frac{ e^{i q r}}{\sqrt{r}}
            \frac{ \sqrt{q}}{ \left| q - q_0 \right| }
            \mu \nonumber \\
           &   & \times \left\{
                            \frac{ \sin \theta}{ 1 - \cos \theta}
                             - e^{ 2 i (q-q_0) r} \frac{\sqrt{q}}{2q_0 - q}
                            \frac{ \sin \theta }
                                 { 1 - \frac{2q( 2q_0 -q)}
                                            { (2q_0 - q)^2 + q^2} \cos \theta
                              }
                        \right\} .
\eeqa We immediately notice that the scattered wave has two parts, one
spherical
wave with momentum $ q $ and one echo with $ 2q_0 - q $. The cross-section will
not only involve the individual contributions of each part but also the
interference. For $ | q -q_0 | \gg \frac{1}{r} $ the interference can be
ignored
and the cross-section will be given by:
\beq
   \sigma (q, \theta) = \pi \left ( \frac{ \mu}{ m} \right) ^2
                        \frac{q}{(q -q_{0} )^2} \cot^2 {\frac{\theta}{2}} .
\label{sroton}
\eeq This is also the same result obtained long ago by Hall and
Vinen\cite{hall}.

\subsection{Superconductor Thin Films}

 Another system where we can use formula (\ref{cs}) is a superconductor
thin film. Scattering cross-section and mutual friction in superconductors were
investigated on the basis of quasi particle-vortex interactions. This is the
dominant contribution for ordinary superconductors. However, at low
temperatures
and small core size, the quasi particles are frozen, and interactions between
vortices and collective excitations become more important.

\p

 In a very broad sense, superconductivity is superfluidity of charged
particles.
The important  distinction is that the vortex singularities are like thin flux
tubes for bulk materials because of the diamagnetic currents set up in the
sample. However, in thin films this screening is only effective beyond a length
$ L_{s} $ given by
\cite{deGennes} :
\beq
   L_{s} = \frac{\lambda_{L} ^ 2}{d} ,
\eeq where $ \lambda_{L} $ is the London penetration depth and $ d $ is the
thickness of the film. This length scale can be quite large compared to the
bulk
value $\lambda_{L}$. Thus the vortex in a thin film is very much like a
vortex in neutral $~^4$He. There is one little difference that is related to
pairing in the ground state \cite{bering}. This forces us to introduce a
modified shift of $ \frac{\hat{\theta}}{2 r} $ in equation~(\ref{sshift}). The
final effect of this on the cross-section is a factor of $\frac{1}{4} $:
\beq
   \sigma (q, \theta) = \frac{\pi}{8} \frac{q}{(\omega'(q))^2}
    \cot^2{\frac{\theta}{2}} .
\eeq

The collective excitations of a superconductor are the density fluctuations of
cooper pairs, and have the same dispersion as the ordinary plasmon at zero
temperature, which has the following form in two dimensions:

\beq
   \omega(q) = \left( \frac{2 \pi n e^2}{m_{e}} \right)^{1/2}
                    q^{1/2} ,    \label{plasmon}
\eeq where $ n $ is the 2D density of electrons, $ e $ the electron charge and
$
m_{e} $ the electron mass. The resulting cross-section is then:
\beq
   \sigma(q, \theta ) = \pi \left( \frac{\hbar}{m_{e} c} \right) ^2
                        L_{s} q^2 \cot^2{\frac{\theta}{2}} , \label{csplasmon}
\eeq where c is the speed of light.

\section{Conclusions} In summary, we have investigated the interaction between
collective excitations and vortices in superfluid systems.
Starting from a Feynman type many-body wave function,
general dynamic equations for the collective excitations in the presence
of a vortex have been derived, in which the specific properties
of a given system enters only through
the density correlation function or the static structure factor.
The scattering cross-section has been derived in the Born
approximation for long wavelengths,
whose angular dependence has the universal form $\cot^2{\frac{\theta}{2}}$,
with a prefactor only involving the energy spectrum of the collective
excitations.
For short wave lengths, the geometrical optics or the WKB approximation to the
dynamical equations has been shown to yield the classical equations of motion
for
wave packets.

\p

We have applied the Born approximation cross-section formula to phonons and
rotons in $~^4$He, and showed that the results are consistent with previous
calculations. As a new result, we have also obtained the cross-section of
plasmons from vortices in superconductor thin films.
These cross-sections, when combined with the formulae in \cite{rayfield}, can
be
used to find the frictional forces on a vortex.
For the case of superconductor thin films,
the longitudinal dissipative force on a moving vortex is shown to be
proportional to a high power of temperature, $ T^ {10} $,
which supports the idea that vortices move
freely at low temperatures.

\p

 The transverse force due to scattering vanishes as
long as the Born approximation is valid, which is the case for phonon and
plasma
scatterings.  However, the dynamics of rotons in superfluid helium
falls rather into the validity range of WKB approximation, and can yield a
transverse force on vortices \cite{hillel,barenghi,sonin}.  This is very
similar to the
scattering of electrons by a magnetic flux line. The classical calculation
valid
at short wavelengths yields a transverse force, whereas the Born approximation
 valid in the long wavelength limit does not.

\p

\section{Acknowledgments}

We would like to thank D. J. Thouless, V. L. Pokrovsky, M. Stone and M.
C. Chang for helpful discussions. This work is supported by the Welch
Foundation.

\end{document}